# Structural phase transition, grain growth and optical properties of uncompensated Ga-V co-doped TiO$_2$

**Nasima Khatun[1], Saurabh Tiwari[2], Jayanti Lal[3], Chuan-Ming Tseng[4], Shun Wei Liu[5], Sajal Biring[6] and Somaditya Sen[1,2*]**

[1]Department of Physics, Indian Institute of Technology Indore, Simrol Campus, Khandwa Road, Indore 453552, India

[2]Metallurgy Engineering and Material Sciences, Indian Institute of Technology Indore, Simrol Campus, Khandwa Road, Indore 453552, India

[3]Department of chemistry, Vels Institute of Science, Technology & Advanced Studies, Chennai, Tamil Nadu – 600117, India

[4]Department of Materials Engineering, Ming Chi University of Technology, New Taipei City - 24301, Taiwan

[5]Organic Electronic Research Center, Ming Chi University of Technology, New Taipei City - 24301, Taiwan

[6]Electronic Engg., Ming Chi University of Technology, New Taipei City -24301, Taiwan

## Abstract

Effect of uncompensated Ga-V co-doping (0≤$x$≤0.046) on structural phase transition, grain growth process and optical properties of TiO$_2$ is reported here. Inhibition of phase transition due to co-doping is confirmed by X-ray diffraction measurement. Activation energy of phase transition increases from 120KJ/mol ($x$=0) to 140 KJ/mol ($x$=0.046) due to Ga-V co-doping. In anatase phase, lattice constants increase by the effect of Ga$^{3+}$ interstitials. This results in inhibition of phase transition. Anatase phase becomes stable up to ~650 °C in co-doped sample whereas for pure TiO$_2$ phase transition starts in between 450-500 $^0$C. In anatase phase, strain increases due to co-doping which reduces crystallite size. In rutile phase, grain growth process is enhanced due to co-doping and particles show a rod-like structure with majority {110} facets. Bandgap decreases in both phases and reduced to a visible light region. BET analysis shows that surface area increases from 4.55 m$^2$/g ($x$=0) to 96.53 m$^2$/g ($x$=0.046) by Ga-V incorporation which provide a large number of active site for photocatalytic activity. Hence, co-doped anatase nanoparticle can be used as a promising candidate for photocatalytic applications using visible light up to a higher temperature ~650 °C.

**Introduction:**

$TiO_2$ is of continual interest due to its multifunctional properties. Different crystal structure and corresponding electronic band structure facilitate its applications in different fields such as in opto-electronic devices[1, 2], self-cleaning glass coating materials[3], photocatalyst [4, 5], fuel cell[6], dye-sensitized solar cell[7, 8], opacifier and white pigment[9, 10], etc.. It facilitates environmentally beneficial reactions through photocatalytic activity by splitting of water to generate hydrogen and treatment of polluted air and water [11]. Low cost, nontoxicity, and high chemical stability add a special importance for application.

$TiO_2$ has three naturally occurring polymorph[12]. In order of abundance, these are rutile (R), anatase (A), and brookite (B)[13]. At lower temperature anatase is the most stable phase due to its low surface free energy [14, 15]. Pure brookite phase is not available at normal ambient condition due to its complex crystal structure. Both anatase and brookite are metastable phases. With increasing temperature (≥ 750 °C) both the phases are irreversibly transformed into stable rutile phase[16]. Density functional theory (DFT) calculation showed that effective mass of electrons and holes is smaller in anatase phase compared to brookite and rutile phases[17]. This facilitates the migration of carriers and enhances photocatalytic activity. Due to lighter effective mass, smaller particle size (highly stable <10-15 nm)[12, 18] and longer lifetime of photogenerated charge carriers[17], anatase is an active polymorph for photocatalytic applications than brookite and rutile[19-21].

In spite of these important properties of anatase $TiO_2$, there are few restraints. Anatase $TiO_2$ has lower thermal stability (≤ 450-500 °C) and wide bandgap (3.2 eV)[22]. Due to its wide bandgap, it only absorbs ~5% radiation of the solar spectrum. The entire visible region (~45%) remains unutilized. From a practical application point of view, utilization of visible light is beneficial. Therefore, people have made significant efforts to shift its phase transition temperature to a higher temperature region and tune the bandgap of $TiO_2$ in visible light range. Different processes have been adapted for this purpose. These include synthesis methods [23-25], strain[26, 27], doping with different elements (Fe, Mo, V, Ru, Cu, Fe, Cr etc.) [14, 28-32], etc.

Among all these processes, doping is the easiest way to control phase transition and thereby properties (tune the bandgap). From the literature, it was observed that Ga doping inhibits the phase transition and play a robust role in photocatalytic activity (PCA) in UV region[33, 34]. Vanadium considerably reduces the bandgap but promotes phase transition [16, 35, 36]. Hence, in this context uncompensated Gallium and Vanadium co-doping have been chosen to overcome both the problem. Here in this work, effect of uncompensated Ga-V co-doping on structural phase transition, grain growth process and optical properties of $TiO_2$ has been discussed.

**Experimental**

Ga and V co-doped $TiO_2$ ($Ti_{(1-x)}(Ga_{0.8}V_{0.2})_xO_2$: TGV) nanoparticles (with $x$=0.00 (TGV0), 0.015 (TGV1), 0.031 (TGV3) and 0.046 (TGV4)) are prepared by modified sol-gel synthesis. The Ti-solution is prepared by mixing required amount of dihydoxy-bis titanium (TALH: $C_6H_{18}N_2O_8Ti$) in deionized (DI) water at room temperature. An appropriate amount of $Ga(NO_3)_3$ is dissolved in DI water in one beaker. In another beaker, $V_2O_5$ is also dissolved in DI water by adding little amount of $NH_4OH$ while stirring. Both the solutions are added dropwise into a Ti-solution. After 1h of mixing, citric acid and ethylene glycol is added to it. This mixture is stirred for another 1h for homogeneous mixing. Thereafter it is slowly heated. Temperature of this solution is maintained at 80 $^0C$ for 4-5h to get the thick gel. The gel is burnt on a hot plate at 100 $^0C$ in normal ambient condition resulting in a black dry powder. This powder is denitrified and decarburized at 450 $^0C$ in an air atmosphere for 6h to get desired nanoparticles. The collected powder is subsequently heated at eight different temperatures in 50 $^0C$ steps from 450 $^0C$ to 800 $^0C$, stabilizing for 6h at each temperature.

Thermal gravimetric analysis (TGA) was performed to estimate the crystallization temperature of the samples by METTLER TOLEDO (TGA/DSC 1) system using the STAR$^e$ software system up to 800 $^0C$ in an air atmosphere with a heating rate of 5 $^0C$ $min^{-1}$. Structural analysis was studied by powder X-ray diffraction (XRD) patterns using Bruker D2 phaser diffractometer with Cu-Kα radiation (λ=1.5418 Å). Morphology and particle size were investigated using high-resolution transmission electron microscope (HRTEM) (JEOL JEM-2100 LaB6, accelerating voltage - 200 kV) and field emission scanning electron microscopy (Supra55 Zeiss- FESEM). Surface area of the samples was calculated by Brunauer–Emmett–Teller (BET) measurement.

Diffuse reflectance spectroscopy (DRS) measurements were carried out using Bentham TMc300 Monochromator to estimate the changes in the bandgap.

**Results and Discussion**

TGA measurement on dry gel powder (TGV0) is performed from room temperature (RT-27 °C) to 800 °C. Weight loss of 1.04%, from RT to ~110 °C, is observed in pure $TiO_2$ (TGV0) (Fig. 1). This is due to elimination of physically adsorbed water[37]. In the temperature regime, ~110 to 315 °C a sharper weight loss of 2.52% is observed. This may be attributed to the rupture of a polymeric chain of black powder and removal of ethylene glycol units[38]. A final very sharp weight loss of 5.61% is observed in the temperature regime ~315 to 430 °C. This drastic loss is due to decomposition of the organic compounds into carbon dioxide and nitrogen dioxide, desorption of chemisorbed water molecules[39, 40]. This temperature is well enough to consume unreacted precursor and thereby form well crystalline samples. Note that beyond ~430-450 °C, there is almost no change in weight loss. Thus, 450 $^0$C is selected as the optimum calcination temperature which is high enough to achieve crystallization, and optimum to reduce the thermal growth of the particles to maintain nano-scale features in the calcined powder. A minor weight loss of 0.48%, in between 550 to 650 °C can be ascribed to the removal of the surface hydroxyls present in the samples[41].

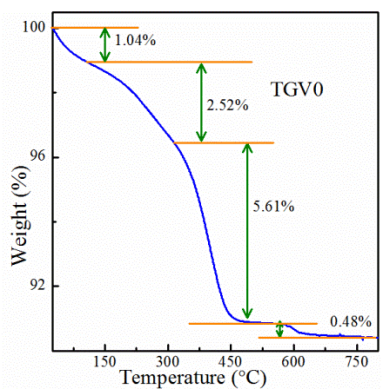

Fig. 1 TGA curve of the TGV0 sample at a temperature range of 27-800 °C.

HRTEM is a very powerful tool to investigate a particle in a very small range (around 1-2 nm). It gives crystallographic information and clear morphology of the nanoparticles. Fig. 2 (a and c) shows the TEM images of TGV0 and TGV3 samples. Almost spherical shape particles are observed for both the samples. Particles size of both the sample has been calculated using Image

J software. Histogram of TGV0 and TGV3 sample (inset of Fig. 2(a) and (c)) shows average particles size are in the range of ~12-15 nm and ~8-10 nm respectively. It is observed that particle size reduces due to co-doping. In V doped $TiO_2$, it was observed that crystallite size reduced with doping [35, 42]. Ga doping also reduces crystallite size[21, 33]. Hence, a combination of V and Ga co-doping is supposed to reduce crystallite size. TEM results confirm the same. In most cases, strain increases upon doping of foreign elements into $TiO_2$ and this strain hinders the grain growth process of nanoparticles. From HRTEM images, it is observed that d spacing of lattice fringes of both the samples are ~0.35 nm which corresponds to 101 planes of anatase $TiO_2$ (Fig. 2 (b: TGV0 and d:TGV3). Clarity of the fringes signifies both the samples are well crystalline. Such a good crystallinity at a low temperature ~450 °C, is possible due to proper choice of specific reagents (ethylene glycol and citric acid)[43, 44] used in this synthesis methods. The ring-like SAED patterns (inset of Fig. 2 (b: TGV0 and d: TGV3) reveals the polycrystalline nature and confirms anatase phase of $TiO_2$ of both the samples.

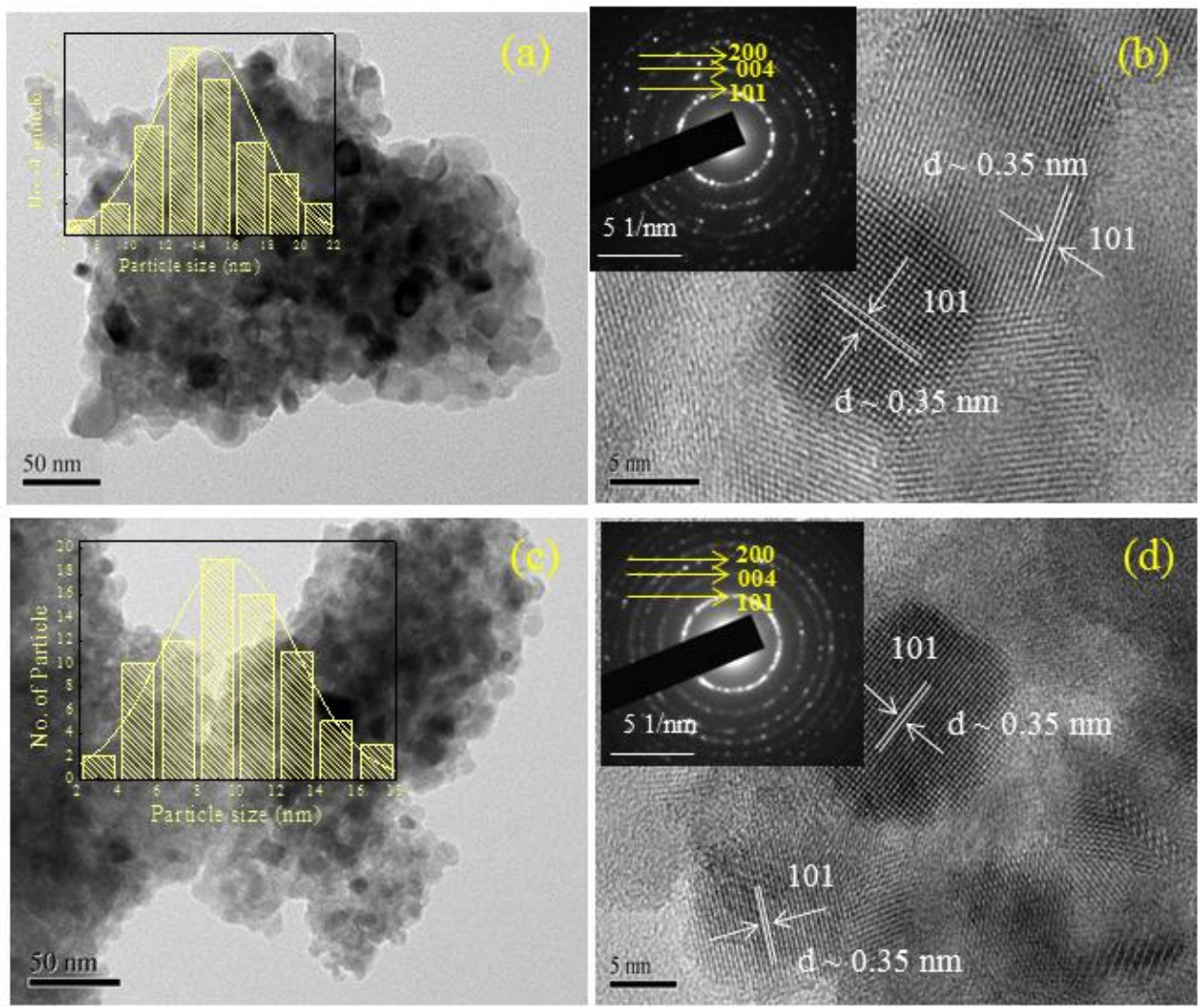

Fig. 2: TEM images of TGV0 (a) and TGV3 (c) and inset shows the histogram of particle size distribution of corresponding samples. (b and d) HRTEM images of TGV0 and TGV3 and insets show SAED pattern of corresponding samples.

Table 1. BET surface area, pore diameter, pore volume of Ga-V co-doped $TiO_2$ samples.

| Sample name | BET surface area ($m^2/g$) | Pore diameter (nm) | Pore volume ($cm^3/g$) |
|---|---|---|---|
| TGV0 | 4.55 | 3.819 | 0.009 |
| TGV1 | 53.95 | 3.823 | 0.044 |
| TGV3 | 85.25 | 3.826 | 0.060 |
| TGV4 | 96.53 | 3.829 | 0.061 |

Nitrogen adsorption/desorption isotherms (Fig. 3) of all the samples display type IV isotherms according to IUPAC classification. Hysteresis loop of the isotherms are of typical H2(a) type[45]. BET surface area increases from 4.55 m$^2$/g (TGV0) to 96.53 m$^2$/g (TGV4). Surface area depends on size and morphology of nanoparticles. Smaller the size, larger is the surface area of nanoparticles. Hence, as BET surface area increases with doping a reduction in crystallite size is expected. Pore size distribution is calculated from BJH method on the desorption isotherms (Fig. 3(e-h)). Pores sizes for all samples are <4 nm. Mesoporous materials have pore diameters ranging from 2-nm to 50nm [45]. Hence, these samples can be classified as mesoporous materials based on the pore diameter and nature of hysteresis loop. A larger surface area due to Ga-V incorporation provides a large number of active sites which makes the materials better for PCA.

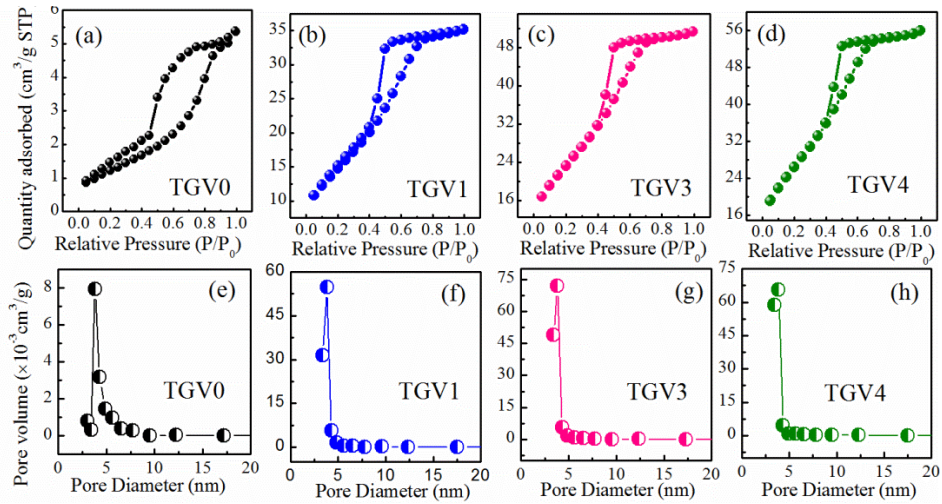

Fig. 3 (a-d) Nitrogen adsorption/desorption isotherms of TGV samples (450 $^0$C) and (e-h) Barrett-Joyner Halenda (BJH) pore size distribution curve of the samples.

XRD pattern for all samples (TGV0, TGV1, TGV3, and TGV4), heated at temperatures ~450 °C, 500 °C, 550 °C, 600 °C, 650 °C, 700 °C, 750 °C, and 800 °C, are shown in Fig. 4. XRD pattern for all TGV samples heated at 450 $^0$C for 6h (Fig. 4(a)) matches well with COD ID-9015929 which is of tetragonal anatase phase of TiO$_2$ having space group $I4_1/amd$. Hence, all the samples are in pure anatase phase. There are no traces of any rutile phase at this temperature. Also, there is no evidence of any simple or complex metal oxide phases related to Ti, Ga, and V. With increasing temperature the anatase phase of TGV samples gradually starts to convert to a rutile phase and forms a mixed phase. Further heating at a higher temperature (~800 $^0$C), all the TGV samples are converted into an entire rutile phase. XRD patterns of TGV samples (800 $^0$C), matches well with COD ID-9009083 which is of tetragonal rutile phase of $TiO_2$ having space group $P4_2/mnm$. The samples heat treated between 450-800 $^0$C shows mixed phase of anatase and rutile.

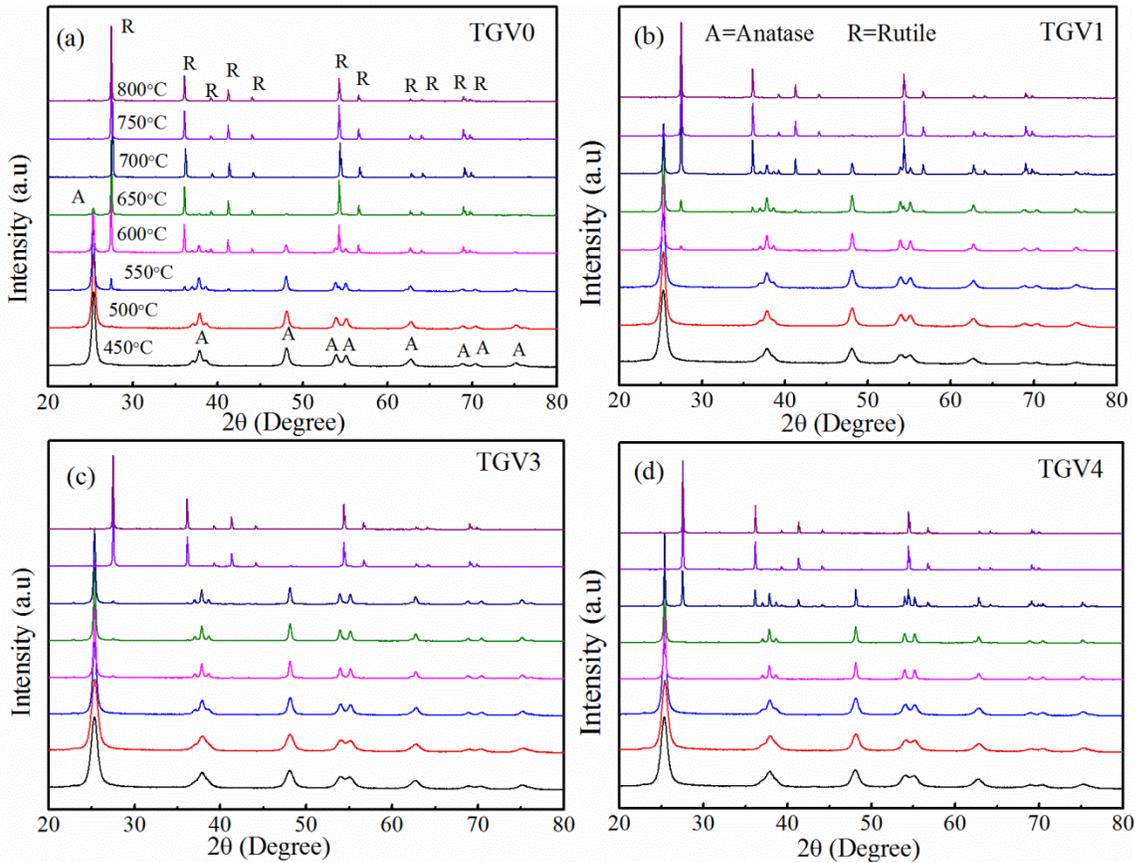

Fig. 4 XRD patterns of all TGV samples at eight different temperatures (450-800 °C) in the range of 2θ=20-80°.

For pure $TiO_2$ (TGV0), A→R phase transformation starts in between ~450-500 °C. Phase transition at this particular temperature happens due to the choice of specific reagent (ethylene glycol and citric acid)[43] used in this method. Complete transformation into rutile phase is observed at ~750 °C. In case of co-doped (Ga-V) samples, no trace of rutile phase has been detected below ~550 °C. For TGV1 and TGV3, A→R phase transition starts ~550-600 °C. For TGV4, A→R phase transition starts at ~650-700 °C. A complete conversion into rutile phase happens at ~800 °C. It is observed from these XRD spectra that the appearance of rutile phase and complete conversion into rutile phase both are shifted to higher temperature with increasing doping concentration. Hence, Ga and V co-doping into $TiO_2$ inhibits the phase transition or stabilize the anatase phase to a higher temperature (for TGV1 and TGV3 up to ~550 °C while for TGV4 up to ~650 °C).

Vigilant investigation on XRD patterns of the samples at rutile phase (800 $^0$C) shows small appearance of $\beta$-$Ga_2O_3$ phase for TGV3 and TGV4 samples (provided in supplementary file Fig. S1) which matches with COD ID-2004987 ($\beta$-$Ga_2O_3$). However, in anatase phase, such type of impurity has not been detected. Anatase phase has some inherent empty space inside crystal structure[46]. Therefore Ga and V easily incorporated into $TiO_2$ lattice and occupy the position of interstitials and substitutional sites. Density ($\rho$) of rutile phase (4.25 gm/cm$^3$) is higher than anatase (3.89 gm/cm$^3$)[14]. Hence, rutile phase has less empty space compared to anatase phase. As $Ga^{3+}$ (0.76Å) ion has slightly bigger ionic radius compared to both $Ti^{4+}$ (0.745Å) and $V^{5+/4+}$ (0.68Å/0.72Å), therefore at higher temperature due to thermal instability and less space, $Ga^{3+}$ ions move out from $TiO_2$ lattice and segregate on the surface of the particles. These Ga ions at higher temperature react with environment oxygen and form $\beta$-$Ga_2O_3$ which are highly dispersed on the surface of particles. $\beta$-$Ga_2O_3$ is a stable crystalline form of gallium oxide at a higher temperature ($\geq$650 °C)[33].

Rutile phase fraction ($f_R$) in the mixed phases is estimated at different temperatures using Spurr and Mayers equation [47]. Temperature dependence of $f_R$ (Fig. 5) for co-doped samples ensures an inhibition of phase transformation with increasing doping concentration. All processing parameters (like heating/cooling rates, environment of calcination, etc.) are kept constant. Hence, this inhibition of phase transformation entirely depends on the concentration of Ga and V co-doping. In general, oxygen vacancy results in lattice contraction and promotes A→R phase transition. On the other hand, interstitials expand the lattice and thereby inhibits the phase transition[48]. Ga ion has slightly bigger ionic (VI-0.76Å) radius and lesser charge +3 compared to $Ti^{4+}$ (VI-0.745Å), while V has variable charge states (3+,4+,5+) with ionic radius ($V^{3+}$ (VI-0.78Å), $V^{4+}$ (VI-0.72Å), and $V^{5+}$ (VI-0.68Å)). From literature, it was observed that charge states and ionic radius are very sensitive to accelerate and delay the A→R phase transition[14]. Ga and V ions have different charge states. Total charge compensation can only happen if amount of Ga and V are equal and the entire V-population is in $V^{5+}$ state. Ga: V ratio in all the samples is 4:1. Hence, for charge compensation, it either creates oxygen vacancies or form interstitials. As discussed above, interstitials are responsible for inhibition of phase transitions and XRD results show inhibition of phase transition due to co-doping. This hints that effect of interstitials is more prominent compared to oxygen vacancies.

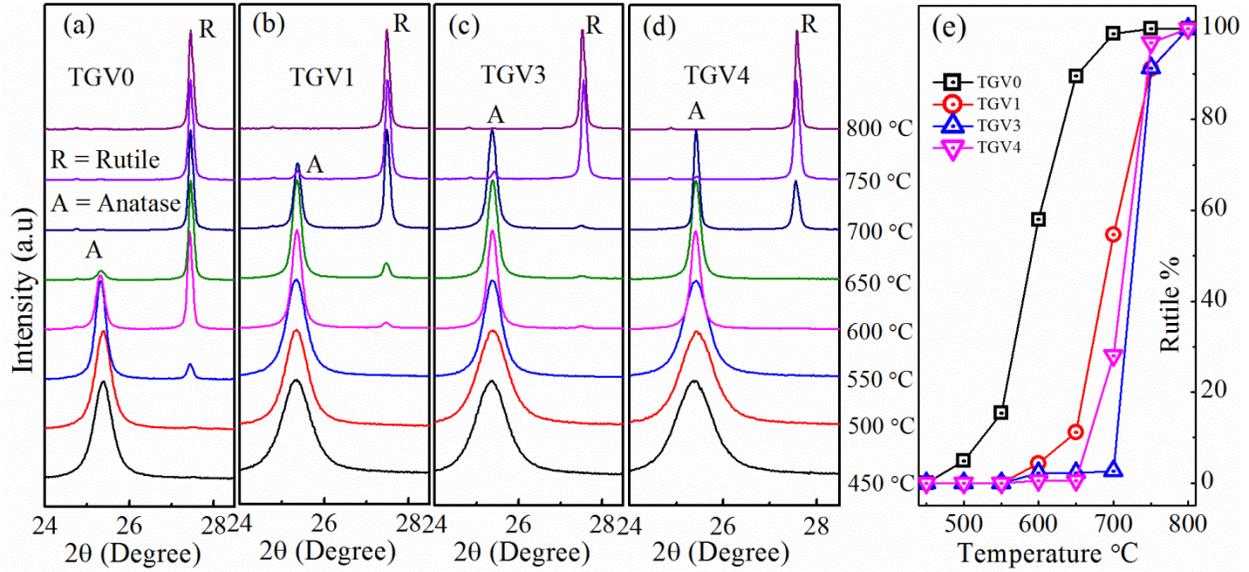

Fig. 5 XRD pattern of all TGV samples at eight different temperature (~450-800 °C) in the range of 2θ=24.5-28°. (b) Fraction of rutile phase ($f_R$) at different temperature.

Activation energy ($E_a$) is the minimum energy required to overcome the energy barrier for A→R phase transition between the two phases. It was also observed from literature[49] that $E_a$ decreases due to oxygen vacancies whereas interstitials are responsible for the increase of $E_a$. $E_a$ is calculated using Arrhenius equation: $ln(f_R) = -\frac{E_a}{RT}$; where, $f_R$ is the fraction of rutile phase present in a sample, R is universal gas constant and T is the temperature in Kelvin. Linear fits of ln($f_R$) vs 1/T gives $E_a$ (Fig. 6(a, b, c, and d)). It is observed that there is a drastic increase in $E_a$ from pure $TiO_2$ (120 KJ/mol) to TGV3 (243 KJ/mol). For TGV4, $E_a$ decreases slightly (240 KJ/mol) from TGV3 but remains higher compared to TGV0 and TGV1. This increasing trend of $E_a$ (Fig. 6(e)) support that the effect of interstitials is more prominent than oxygen vacancies which expands the lattice in Ga-V co-doped samples. This expansion of lattice results in inhibition of phase transition and is consistent with XRD results.

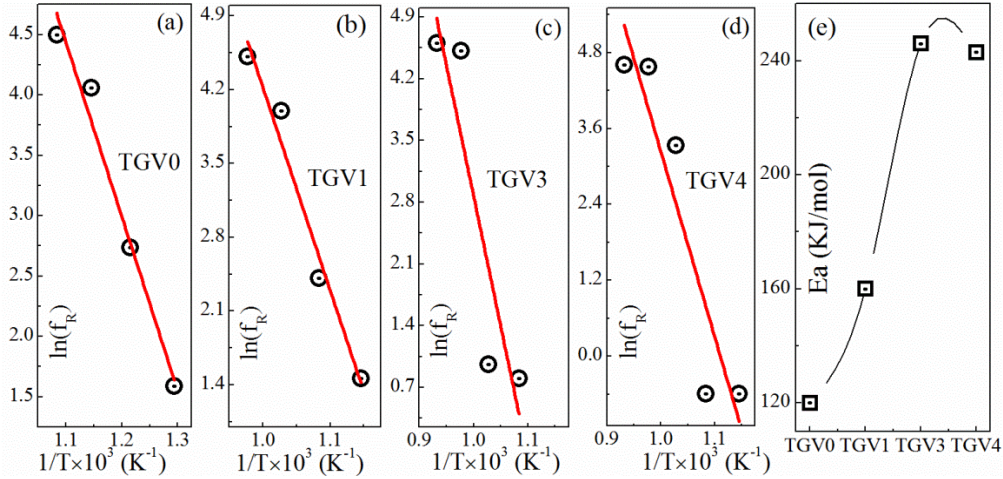

Fig. 6 Fits of ln($f_R$) vs 1/T ((a): TGV0; (b): TGV1; (c): TGV3 and (d): TGV4). (b) Variation of activation energy with doping concentration (solid line is just a guide to the eye).

Samples in anatase phase, when heated to a higher temperature (≥500 °C) leads to rearrangement of Ti-O bonds as a result unit cell volume contracts and phase transformation (A→R) occurs. In anatase phase, lattice constant 'a=b' (3.785 Å) is smaller and 'c' (9.514 Å) is larger compared to lattice constants of rutile phase (a=b=4.59 4Å and c=2.958 Å)[14]. Hence, unit cell volume of anatase phase is larger (136.3 Å$^3$) compared to rutile phase (62.4 Å$^3$). For phase transition, 'a' always increase and 'c' decrease. Hence, this delay of phase transition can be explained in terms of change in lattice constants.

Fig. 7(a) shows the Rietveld refinement of TGV samples in pure anatase phase (450 °C). It is observed that all the three lattice constant increases with increasing doping concentration (Fig. 7(b)). Unit cell volume also follows the similar trend as observed in lattice constants (Fig. 7(c)). $Cr^{3+}$ has a comparable ionic radius (0.755 Å) as $Ga^{3+}$ and has same charge state. Zhu *et al.*[50] from their DFT calculation showed that anatase phase formation energy is low when $Cr^{3+}$ occupies interstitials sites than substitutional sites. With increasing doping concentration, $Cr^{3+}$ going from interstitial to substitutional sites was observed to vary. Hence, $Ga^{3+}$ ions too may have the same tendency to go more into interstitial sites than substitutional sites. Banerjee et al.[51] experimentally showed that $Ga^{3+}$ ions occupy more interstitial sites than substitutional sites in $TiO_2$. These interstitial sites are responsible for the expansion of lattice and inhibit the phase transition. Depero *et al.*[34] experimentally proved that Ga doping inhibits phase transition. It was also observed that formation energy of anatase $TiO_2$ is low when V occupies the substitutional sites rather than interstitial sites[52]. Hence, theoretically and experimentally it was proved that V occupies substitutional sites in $TiO_2$ [35, 53, 54] and thereby decrease all the three lattice constants. This is because $V^{4+/5+}$ have smaller ionic radius compared to $Ti^{4+}$. Vittadini et al.[55] reported that $V^{5+}$ is more likely the major surface species where $V^{4+}$ is stable inside balk. From TEM results it was observed that particle is in a spherical shape and in nano size. With increasing doping concentration, particle size decreases which result to increase the surface area to volume ratio of the samples. BET measurement shows surface area increases with increasing doping concentration. Hence at anatase phase, all the Vanadium ions are mostly in 5+ oxidation states[42, 54]. Hence, contraction of lattice constants and thereby unit cell volume by V incorporation promoted the A→R phase transition[16]. In all the co-doped samples as Ga content is more compared to V (Ga:V=4:1), hence the effect of Ga interstitial play a significant role over V substitution and oxygen vacancies which expands the lattice. Rietveld refinement on anatase phase shows this expansion of lattice and results in inhibition of phase transition.

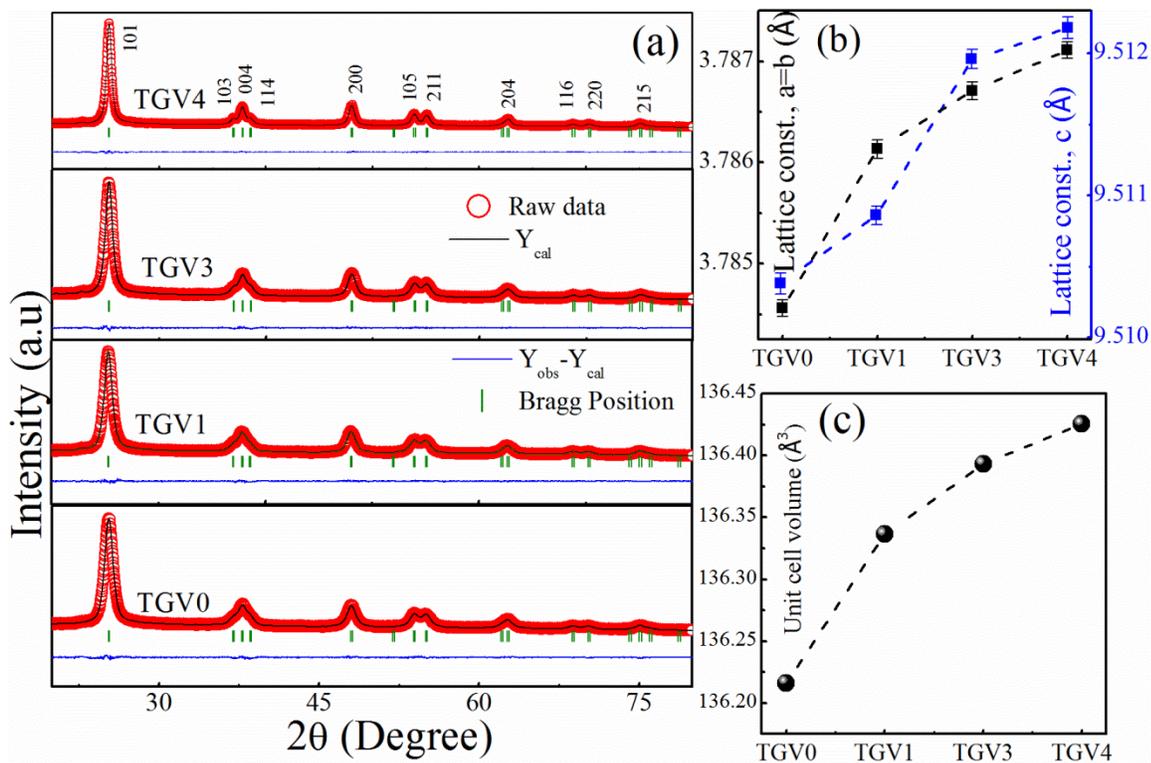

Fig. 7(a) Rietveld refinement of anatase TGV samples (450 °C). Change of lattice constants (b), and unit cell volume (c) with Ga-V doping concentration.

From Rietveld refinement on rutile phase (800 $^0$C), it is observed that lattice constant 'a' and 'b' increases with increasing doping concentration. However, lattice constant 'c' nominally increases for TGV1 and thereafter decreases rapidly for TGV3 and TGV4. Unit cell volume also increases for TGV1 and thereafter decreases (for TGV3 and TGV4). As mentioned above that for TGV3 and TGV4, due to thermal instability and less space, few Ga ions move out from lattice structure. Hence, relative percentage of $V^{4+/5+}$ ions compared to $Ga^{3+}$ ions increases from targeted values (4:1). At rutile phase, due to high temperature (800 $^0$C) particle size increases for all TGV samples and with increasing doping content grain growth process enhanced (discussed later at Fig. 9). Hence surface area to volume ratio decreases which results to increase in $V^{4+}$ species in the samples as discussed above that $V^{4+}$ is more likely stable into bulk. In our earlier report[16], it was observed that V ions are in mixed valence states of $V^{5+}/V^{4+}$ and with increasing doping concentration presence of $V^{4+}$ ions increases. In rutile phase, both the $V^{5+}$ and $V^{4+}$ ions occupy the substitutional sites in $TiO_2$ lattice. $V^{5+/4+}$ ions have smaller ionic radius compared to $Ti^{4+}$ and $Ga^{3+}$ which results in a decrease of lattice constants 'c' as well as unit cell volume. $Ga^{3+}$ ions occupy more interstitials sites than substitutional sites in $TiO_2$ discussed above. At lower doping (TGV1), as all the Ga ions are inside the crystal structure and due to significant role of this $Ga^{3+}$ interstitials unit cell volume increased. However at higher doping (TGV3 and TGV4), due to substitutional $V^{4+/5+}$ ions and oxygen vacancies unit cell volume decreased.

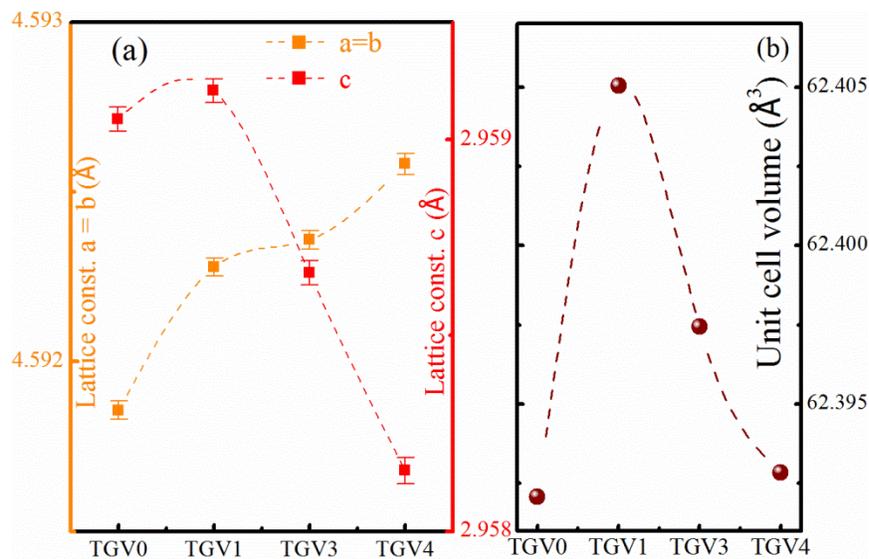

Fig. 8 Change in lattice constants (a) and unit cell volume (b) with co-doping concentration at rutile phase (800 °C).

To investigate the effect of Ga and V co-doping on grain growth process, crystallite size (at 450 °C) is calculated using Scherrer equation. It is observed that crystallite size decreases from 14.5 (pure $TiO_2$) nm to 8.9 nm (TGV4) by Ga-V incorporation (Fig. 9(e)). Hence co-doping restrains the grain growth process of anatase nanoparticles which is consistent with TEM results. In most metal oxides this restrains grain growth is due to increasing strain in the nanoparticles.

Such increase of strain in lattice due to Ga-V co-doping has been verified by the shape and peak positions of pure anatase TGV samples (at 450 °C). Usually, crystallites in polycrystalline aggregates are in a state of compression or tension by its neighboring crystallites which produce uniform or non-uniform strain in the lattice. From literature, it was observed that shifting of diffraction peak creates uniform strain whereas peak broadening without changing peak position creates non-uniform strain [51]. Careful inspection reveals that the 101 peak become broad but position remain almost same in TGV samples (provided in supplementary file Fig. S2). Hence, incorporation of $Ga^{3+}/V^{4+/5+}$ at lattice site and as well as in interstitial site may be responsible for such nonuniform stain. Williamson Hall plot is used to calculate quantitative changes in strain due to Ga-V co-doping. Slop of linear fits of $\beta\cos(\theta)$ vs $4\sin(\theta)$ gives strain; where $\beta$ is the FWHM of corresponding peaks of XRD spectra. It is observed that strain increases with doping concentration (Fig. 9(e)). This increasing strain due to Ga-V incorporation retards the grain growth process of anatase nanoparticle.

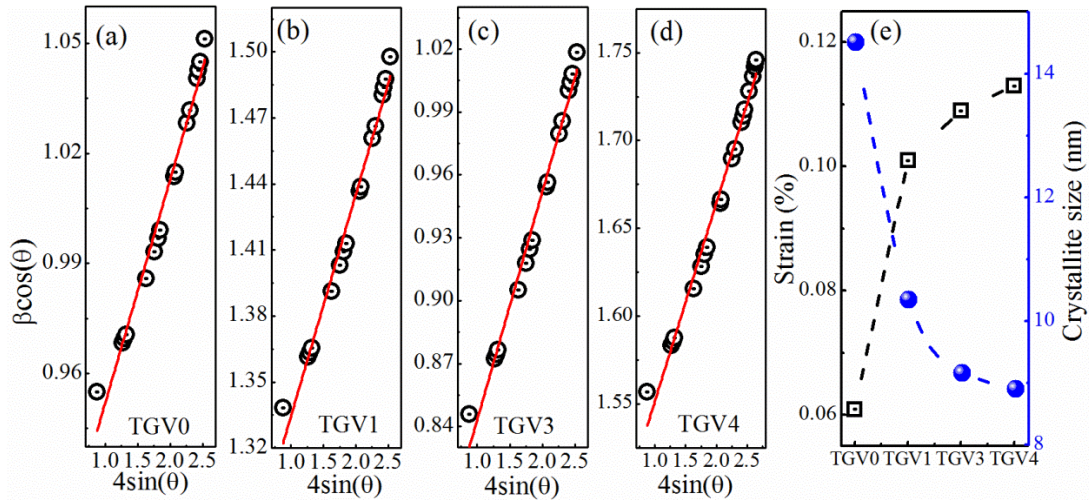

Fig. 9 Linear fits of $\beta\cos(\theta)$ vs $4\sin(\theta)$ of TGV samples in anatase phase (450 °C). (e) change of strain and crystallite size with doping concentration.

Fig. 10 shows the FESEM images of TGV samples at rutile phase (800 $^0$C). Oliver et al.[56], from their DFT calculations on rutile phase, reported that {110} surface has the lowest surface energy (1.78 J/m$^2$) whereas {100} surface perpendicular to {110} surface has the highest surface energy (2.08 J/m$^2$). During crystal growth, the low energy surface {110} grows fastest and high energy surface {100} tend to decrease its surface area to minimize the total energy per crystal[57]. As mentioned above, particles of anatase TGV samples (450 $^0$C) are almost in a spherical shape. With increasing temperature anatase phase converted into mixed phase and with further heating transform into an entirely rutile phase. Similarly, particle shape and size also changed with temperature. With increasing temperature spherical anatase crystals enlarge its size and become elongate spherical to a rod-like structure. For pure TiO$_2$ (TGV0), particles are in irregular spherical shape or distorted rod-like structure and average particle size is ~200 nm (calculated using Image J software). From these images (all the images are in same magnification), it clearly observed that particles have prominent rod-like structure and grain growth process enhanced with increasing doping concentration. It was observed that Ga doping restrain the rutile grain growth process[34]. Whereas in earlier reports it was observed that V enhanced the grain growth process[16]. Effect of Vanadium is more sensitive in grain growth process of rutile particle than Gallium. Hence, the faster grain growth process in co-doped samples is mainly due to the effect of vanadium. According to the nature of surface edges and area at par with reported literature, {110} surfaces are the most prominent surfaces of rutile particle[56, 58, 59]. Some portions of FESEM images of TGV samples have been zoomed to show these surfaces. For TGV0 samples, the zoomed view is shown in the inset of Fig. 10(a). The zoomed views of the co-doped samples are shown as ② (TGV1), ③ (TGV3), and ④ (TGV4).

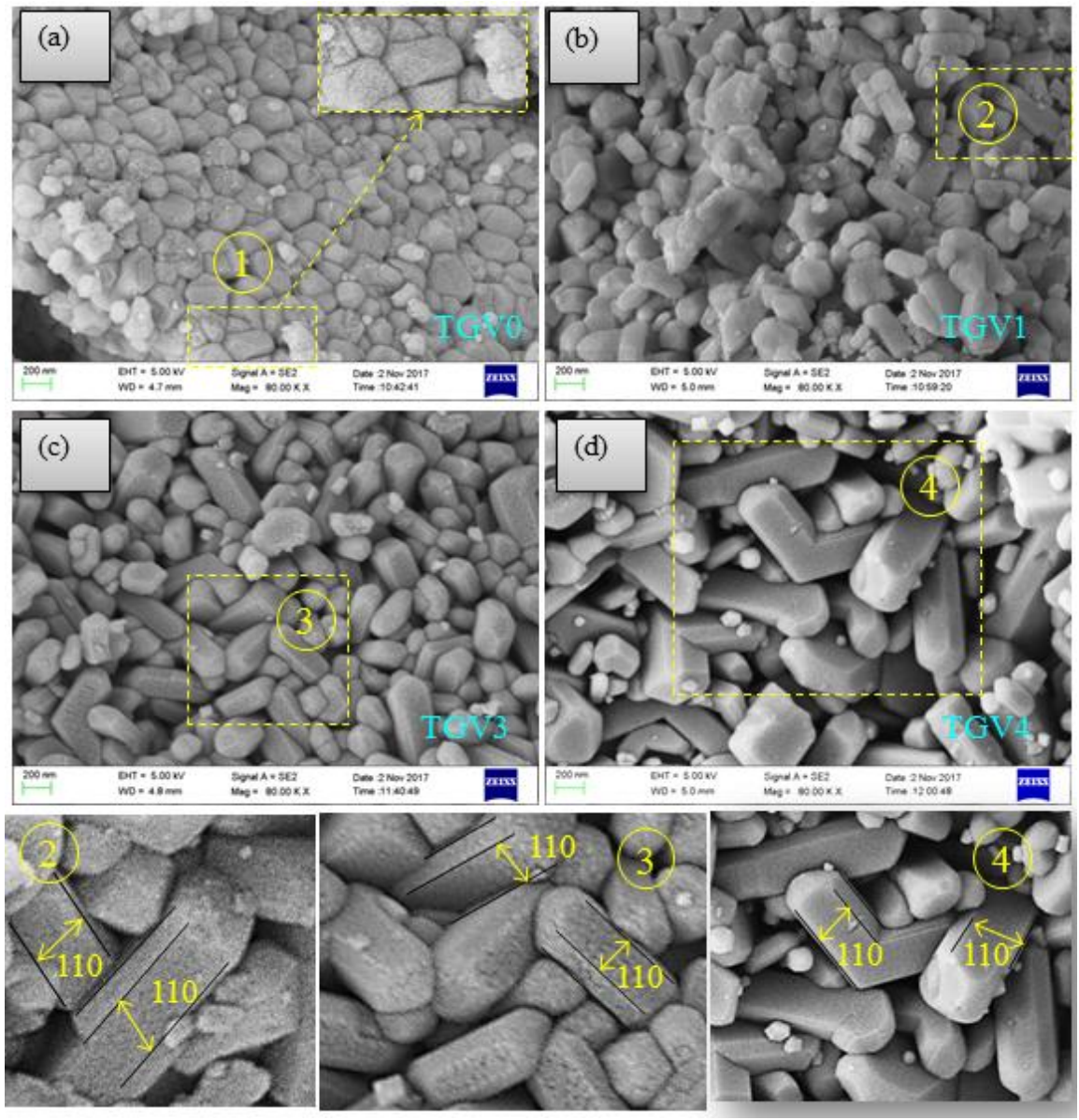

Fig. 10 FESEM images of TGV samples ((a): TGV0; (b): TGV1; (c): TGV3 and (d): TGV4) heated at 800 $^0$C.

TEM images of TGV1 sample heated at 800 $^0$C are shown in Fig. 11((a) and (c)). SAED patterns of the particle are shown as insets of corresponding images and it belongs to the rutile phase of $TiO_2$. The d-spacing of HRTEM images are also belonging to the same crystallographic planes (particle #1, $d_{110}$~0.33 nm and $d_{101}$~0.25nm (Fig. 11(b)) while in particle #2, $d_{110}$~0.33 nm and $d_{001}$~0.29 nm (Fig. 11 (d))). From these images, it observed that {110} facets are the major surface of the particle. The size of the particles seems to be in the range of ~ 100 nm and beyond.

This is smaller than the average size obtained from FESEM studies. A possible reason may be the sample preparation process for TEM measurements. Disperse solution of samples are prepared in ethyl alcohol and a droplet is dropped on TEM grids and dried. In most cases, only the lighter and smaller particles get selected in this process.

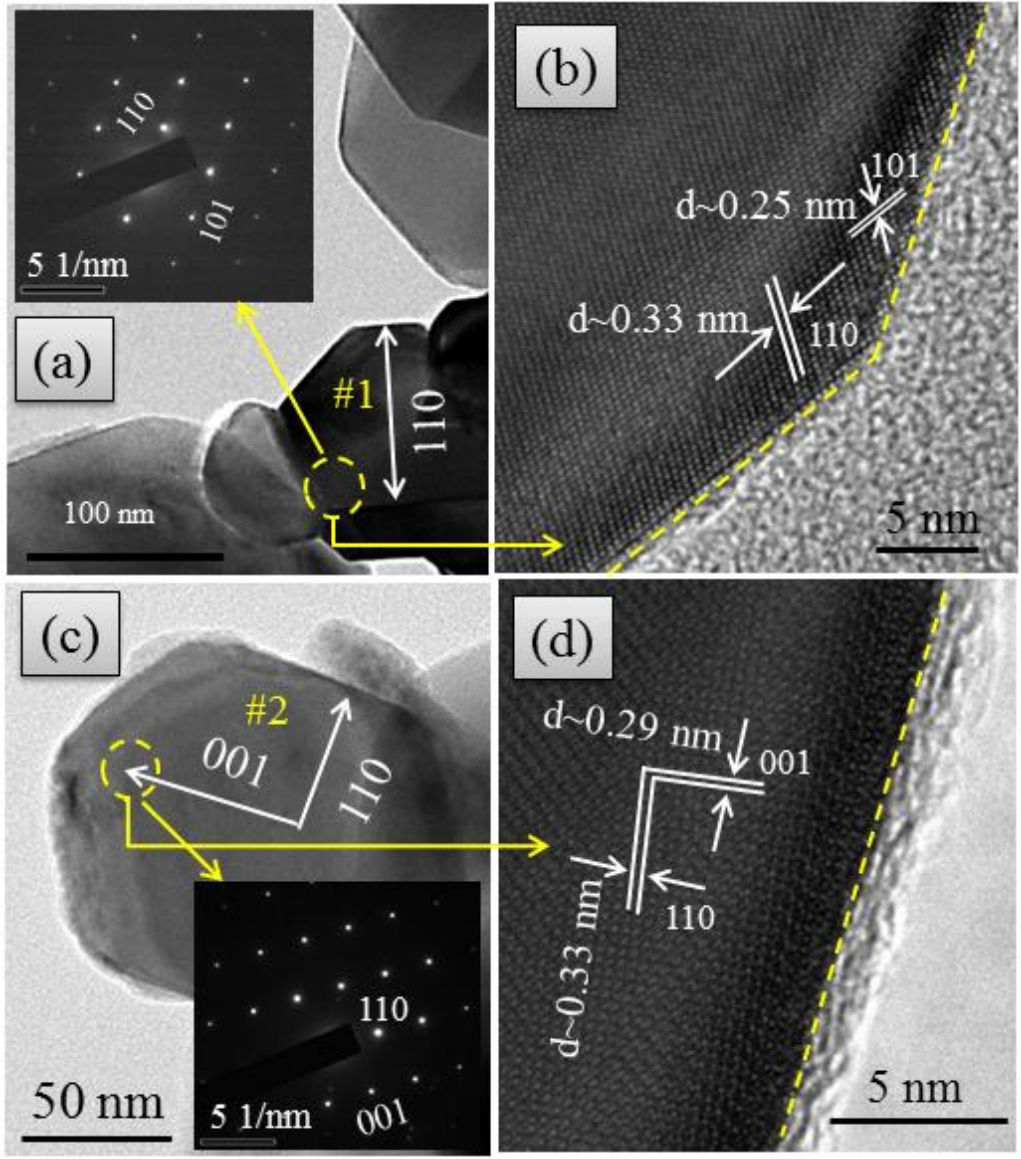

Fig. 11 TEM image of TGV1 sample ((a) and (c)) heated at 800 $^0$C and inset show the corresponding SAED patterns. HRTEM images ((b) and (d)) show lattice fringes of the particle.

Ga-V co-doping also affects the optical properties of $TiO_2$. Room temperature DRS measurement has been carried out to investigate the bandgap of the samples (Fig. 12(a) and (b)). Bandgap is calculated using Tauc plot $(F(R))h\upsilon = A(h\upsilon - E_g)^n$, where, R is the reflectance, A is a constant, h is the frequency of illumination, $E_g$ is the bandgap and n is a unitless parameter with value ½ or 2 for direct or indirect bandgap semiconductor respectively. From literature, it was observed that anatase is an indirect (n=2) bandgap whereas rutile is direct (n=1/2) bandgap semiconductor[17]. It is observed that bandgap decreases due to co-doping for both the phases. At anatase phase, bandgap decreases from 3.14 eV (TGV0) to 2.86 eV (TGV4) (Fig. 12(c)) and at rutile phase, bandgap decreases from 3.06 (TGV0) to 2.84 eV (TGV4) (Fig. 12(d)). O 2p and Ti3d hybridization (p-d) form strong bonding states which are responsible to form valence band (VB) in $TiO_2$. On the other hand, antibonding states due to p-d hybridization between O 2p, Ti 3d, and Ti 2p form conduction band (CB)[17]. Due to Ga doping, the hybridization between O 2p, Ti 3d, and Ga 3d becomes weaker and form empty states in the bandgap of $TiO_2$ which therefore widen the conduction band[33, 60]. This results in increasing the bandgap. From literature, it was also observed that effect of Ga for enhancement of bandgap is not much pronounced[21, 33, 51, 61]. In case of V doped $TiO_2$, p-d hybridization of O 2p, Ti 3d, and V 3d form impurity energy levels (or donor levels) inside the bandgap. Due to these energy levels bandgap decreases[42]. It was also observed that effect of V is more sensitive compared to Ga for bandgap change. Hence, due to combined effect of both V and Ga, bandgap decreases in co-doped samples.

In case of rutile phase, bandgap gradually shifted to lower values from 3.06 eV (TGV0) to 2.84 eV (TGV4) with increasing doping concentration. For TGV1, bandgap shows slight red shift because V concentration is very low in the samples (Ga: V ~ 4:1). For TGV3 and TGV4, it is observed from XRD data that some amount of $Ga^{3+}$ ions move out of the lattice and forms β-$Ga_2O_3$ phase. Hence, relative amount of $V^{4+/5+}$ to $Ga^{3+}$ is increased (Ga:V=4:1; ratio decreased) compared to intended values. It was discussed that $V^{4+}$ ion is more effective than $V^{5+}$ ion in reducing the bandgap[62] of $TiO_2$: ($E_g(V^{4+}) < E_g(V^{5+})$). This may also be a reason, for a gradual shift of bandgap in rutile phase.

Structural modifications are inevitable when foreign elements are incorporated into any lattice. It is known that electronic band structure is strongly correlated with a lattice structure. Urbach energy ($E_U$) is a measure of lattice distortion in the samples which affects electronic band structure [63, 64]. In most semiconductors, it was observed that bandgap decreases if $E_U$ increases[65, 66]. $E_U$ is calculated from linear fits of "lnF(R)-hυ" plots just below the absorption edge of DRS data. Reciprocal of the slope gives $E_U$[54]. In anatase phase, $E_U$ decrease from 145 meV for TGV0 to 506 meV for TGV4 (Fig. 12(c)). Whereas for rutile phase; $E_U$ decrease from 63 meV for TGV0 to 256 meV for TGV4 (Fig. 12(d)). This increase in $E_U$ signifies a more distorted lattice due to co-doping and a band tailing (Urbach tail) just below absorption edge.

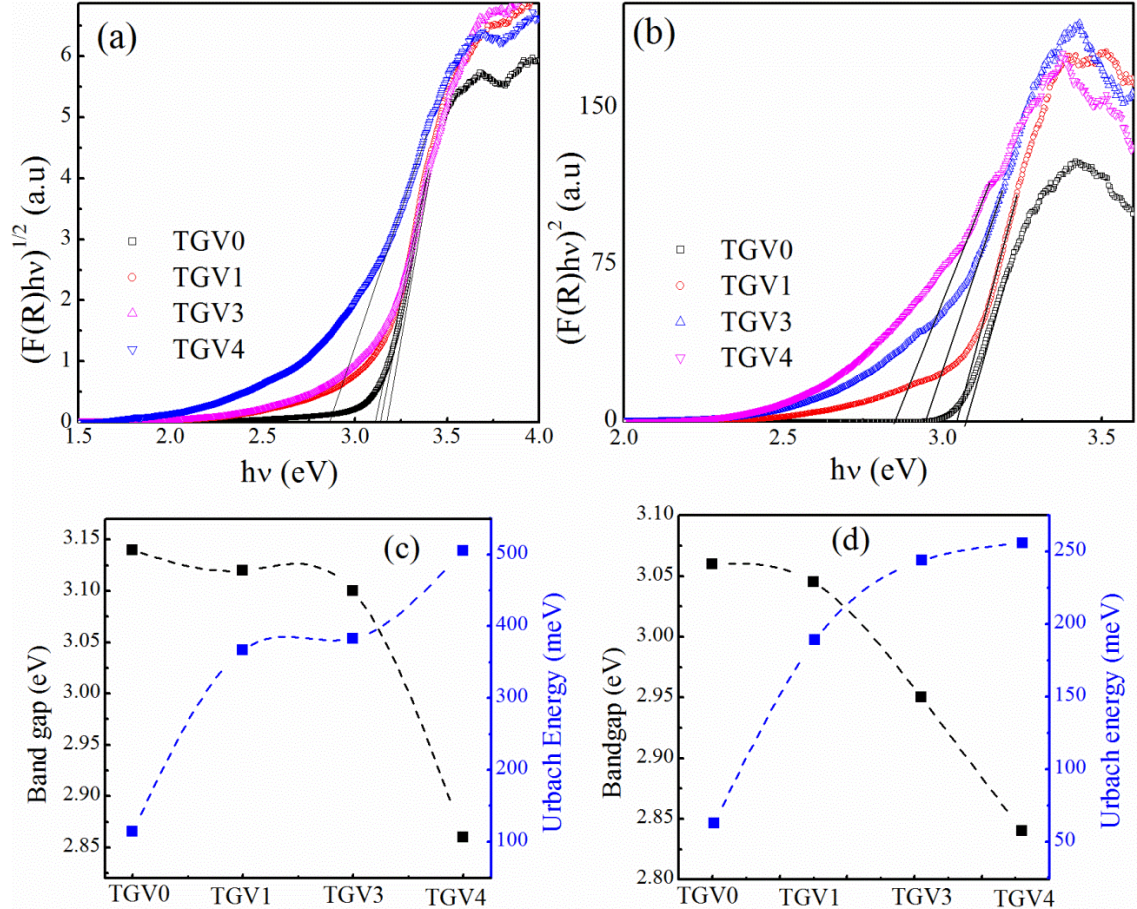

Fig. 12 Room temperature DRS data of co-doped (Ga-V) samples in anatase (a) and rutile phase (b). Inset shows the change of bandgap with co-doping concentration for corresponding samples.

As mentioned above for pure $TiO_2$, phase transition (A→R) starts in between 450-500 $^0C$. Upon co-doping, anatase phase becomes stable up to ~650 $^0C$. Surface area plays an important role for PCA. In the anatase phase, nano-sized spherical particles (8-15 nm) provide a large surface area and BET analysis confirms this increasing trend of surface area with doping concentration. This provides a larger number of active sites for PCA. V decreases the bandgap to visible regions while Ga inhibits phase transition, thereby making the materials a promising candidate for "high temperature visible light photocatalytic application."

## Conclusions

Uncompensated Ga-V co-doped $TiO_2$ has been prepared successfully by modified sol-gel process. Inhibition of phase transition due to co-doping is confirmed by XRD spectra. Activation energy of phase transition increases from 120 KJ/mol ($x=0$) to 240 KJ/mol ($x=0.046$) by Ga-V incorporation which also reveals this inhibition of phase transition. Ga ions occupy more interstitial sites than substitutional sites whereas V ions occupy substitutional sites in $TiO_2$. In anatase phase, lattice constant increases by the effect of $Ga^{3+}$ interstitials, as Ga content is more than V content (Ga:V:=4:1). This expansion of lattice results in inhibition of phase transition. Hence, anatase phase becomes stable up to ~650 °C in co-doped sample. In anatase phase, grain growth process restrained as strain increases by the effect of co-doping and thereby reduces crystallite size. In rutile phase, grain growth process in co-doped sample is enhanced mainly by the effect of Vanadium. Bandgap decreases in both phases and reduces to the visible light region. BET analysis shows that surface area increases from 4.55 $m^2/g$ ($x=0$) to 96.53 $m^2/g$ ($x=0.046$) by Ga-V incorporation. Hence, co-doped anatase nanoparticles can be used as a good photocatalyst using visible light up to a higher temperature ~650 °C.

## Acknowledgment

The authors are sincerely thanking IIT Indore for providing funds and all research related facilities. The authors also thank Sophisticated Instrument Centre (SIC) of IIT Indore for TGA, BET, and FESEM studies. One of the authors (Dr. Sajal Biring) acknowledges support from Ministry of Science and Technology, Taiwan (MOST 105-2218-E-131-003 and 106-2221-E-131-027).

## References

1. M. Zukalova, A. Zukal, L. Kavan, M. K. Nazeeruddin, P. Liska and M. Gratzel, Nano. Lett. **5** (9), 1789-1792 (2005).
2. B. E. Hardin, E. T. Hoke, P. B. Armstrong, J.-H. Yum, P. Comte, T. Torres, J. M. J. Fréchet, M. K. Nazeeruddin, M. Grätzel and M. D. McGehee, Nat. Photonics **3** (7), 406-411 (2009).
3. N. Abidi, L. Cabrales and E. Hequet, ACS Appl. Mater. Interfaces **1** (10), 2141-2146 (2009).
4. C. Fan, C. Chen, J. Wang, X. Fu, Z. Ren, G. Qian and Z. Wang, Sci. Rep. **5**, 11712 (2015).
5. R. Ren, Z. Wen, S. Cui, Y. Hou, X. Guo and J. Chen, Sci. Rep. **5**, 10714 (2015).
6. B. Park and E. J. Cairns, Electrochem. Comm. **13** (1), 75-77 (2011).
7. T. K. Das, P. Ilaiyaraja, P. S. V. Mocherla, G. M. Bhalerao and C. Sudakar, Sol. Energy Mater. Sol. Cells **144**, 194-209 (2016).
8. J.-Y. Liao, J.-W. He, H. Xu, D.-B. Kuang and C.-Y. Su, J. Mater. Chem. **22** (16), 7910 (2012).

9. M. J. A. Ruszala, N. A. Rowson, L. M. Grover and R. A. Choudhery, Int. J. Chem. Eng. Appl. **6** (5), 331-340 (2015).
10. Y. Wang, J. Li, L. Wang, T. Qi, D. Chen and W. Wang, Chem. Eng.Technol. **34** (6), 905-913 (2011).
11. T. Kamegawa, J. Sonoda, K. Sugimura, K. Mori and H. Yamashita, J. Alloys. Compound. **486** (1-2), 685-688 (2009).
12. V. N. Koparde and P. T. Cummings, ACS Nano **2** (8), 1620-1624 (2008).
13. J. Muscat, V. Swamy and N. M. Harrison, Phys. Rev. B **65** (22) (2002).
14. D. A. H. Hanaor and C. C. Sorrell, J. Mater. Sci. **46** (4), 855-874 (2010).
15. K. Ding, Z. Miao, B. Hu, G. An, Z. Sun, B. Han and Z. Liu, Langmuir **26** (12), 10294-10302 (2010).
16. N. Khatun, Anita, P. Rajput, D. Bhattacharya, S. N. Jha, S. Biring and S. Sen, Ceram. Int. **43** (16), 14128-14134 (2017).
17. J. Zhang, P. Zhou, J. Liu and J. Yu, Phys. Chem. Chem. Phys. **16** (38), 20382-20386 (2014).
18. N. Satoh, T. Nakashima and K. Yamamoto, Sci. Rep. **3**, 1959 (2013).
19. Q. L. XiaoBo Li, XiaoYing Jiang, Jianhua Huang, Int. J. Electrochem. Sci. **7**, 11519-11527 (2012).
20. A. Sclafani and J. M. Herrmann, J. Phys. Chem. **100** (32), 13655-13661 (1996).
21. A. K. Chandiran, F. d. r. Sauvage, L. Etgar and M. Graetzel, J. Phys. Chem. C **115** (18), 9232-9240 (2011).
22. B. Buchholcz, H. Haspel, Á. Kukovecz and Z. Kónya, CrystEngComm **16** (32), 7486 (2014).
23. C. Huang, X. Liu, L. Kong, W. Lan, Q. Su and Y. Wang, Appl. Phys. A **87** (4), 781-786 (2007).
24. A. Zaleska, Recent Patents on Engineering **2** (3), 157-164 (2008).
25. C. Rath, P. Mohanty, A. C. Pandey and N. C. Mishra, J. Phys. D: Appl. Phys. **42** (20), 205101 (2009).
26. G. Rajender and P. K. Giri, J. Alloys. Compound. **676**, 591-600 (2016).
27. Anita, A. K. Yadav, N. Khatun, S. Kumar, C.-M. Tseng, S. Biring and S. Sen, J. Mater. Sci.: Mater. Electron. (2017).
28. W. Choi, A. Termin and M. R. Hoffmann, J. Phys. Chem. **98** (51), 13669-13679 (1994).
29. B. Choudhury, A. Choudhury and D. Borah, J. Alloys. Compound. **646**, 692-698 (2015).
30. Y. Zhang, Y. Shen, F. Gu, M. Wu, Y. Xie and J. Zhang, Appl. Surf. Sci. **256** (1), 85-89 (2009).
31. N. Khatun, R. Amin, Anita and S. Sen, AIP Conf. Proc. **1953**, 040028 (2018).
32. G. Hassnain Jaffari, A. Tahir, N. Z. Ali, A. Ali and U. S. Qurashi, J. Appl. Phys. **123** (16), 161541 (2018).
33. S. Luo, T. D. Nguyen-Phan, D. Vovchok, I. Waluyo, R. M. Palomino, A. D. Gamalski, L. Barrio, W. Xu, D. E. Polyansky, J. A. Rodriguez and S. D. Senanayake, Phys. Chem. Chem. Phys. **20** (3), 2104-2112 (2018).
34. L. E. Depero, A. Marino, B. Allieri, E. Bontempi, L. Sangaletti, C. Casale and M. Notaro, J. Mater. Res. **15** (10), 2080-2086 (2011).
35. W. Avansi, R. Arenal, V. R. de Mendonça, C. Ribeiro and E. Longo, CrystEngComm **16** (23), 5021 (2014).
36. S. Klosek and D. Raftery, J. Phys. Chem. B **105** (14), 2815-2819 (2001).
37. S. Shang, X. Jiao and D. Chen, ACS Appl. Mater. Interfaces **4** (2), 860-865 (2012).
38. X. Jiang, T. Herricks and Y. Xia, Adv. Mater. **15** (14), 1205-1209 (2003).
39. H. Yaghoubi, N. Taghavinia, E. K. Alamdari and A. A. Volinsky, ACS Appl. Mater. Interfaces **2** (9), 2629-2636 (2010).
40. M. V. Swapna and K. R. Haridas, J. Exp. Nanosci. **11** (7), 540-549 (2015).
41. V. Caratto, L. Setti, S. Campodonico, M. M. Carnasciali, R. Botter and M. Ferretti, J. Sol-Gel Sci. Technol. **63** (1), 16-22 (2012).
42. W. Zhou, Q. Liu, Z. Zhu and J. Zhang, J. Phys. D: Appl. Phys. **43** (3), 035301 (2010).

43. K. Farhadian Azizi and M. M. Bagheri-Mohagheghi, J. Sol-Gel Sci. Technol. **65** (3), 329-335 (2012).
44. A. B.-N. A. Shalaby, R. Iordanova, Y. Dimitriev, J. Chem. Technol. Metall. **48**, 585-590 (2013).
45. M. Thommes, K. Kaneko, A. V. Neimark, J. P. Olivier, F. Rodriguez-Reinoso, J. Rouquerol and K. S. W. Sing, Pure Appl. Chem. **87** (9-10) (2015).
46. W.-J. Yin, S. Chen, J.-H. Yang, X.-G. Gong, Y. Yan and S.-H. Wei, Appl. Phys. Lett. **96** (22), 221901 (2010).
47. R. A. Spurr and H. Myers, Anal.l Chem. **29** (5), 760-762 (1957).
48. R. D. Shannon and J. A. Pask, J. Am. Ceram. Soc. **48** (8), 391-398 (1965).
49. M. A. Malati and W. K. Wong, Surf. Technol. **22** (4), 305-322 (1984).
50. W. Zhu, X. Qiu, V. Iancu, X. Q. Chen, H. Pan, W. Wang, N. M. Dimitrijevic, T. Rajh, H. M. Meyer, 3rd, M. P. Paranthaman, G. M. Stocks, H. H. Weitering, B. Gu, G. Eres and Z. Zhang, Phys. Rev. Lett. **103** (22), 226401 (2009).
51. A. N. Banerjee, S. W. Joo and B.-K. Min, J. Nanomater. **2012**, 1-14 (2012).
52. R. Long and N. J. English, Chem. Phys. Chem. **11** (12), 2606-2611 (2010).
53. M. Amini, H. Naslhajian and S. M. F. Farnia, New J. Chem. **38** (4), 1581 (2014).
54. N. Khatun, E. G. Rini, P. Shirage, P. Rajput, S. N. Jha and S. Sen, Mater. Sci. Semicon. Proc. **50**, 7-13 (2016).
55. A. Vittadini, M. Casarin and A. Selloni, J Phys Chem B **109** (5), 1652-1655 (2005).
56. P. M. Oliver, G. W. Watson, E. T. Kelsey and S. C. Parker, J. Mater. Chem. **7** (3), 563-568 (1997).
57. A. S. Barnard and P. Zapol, Phys. Rev. B **70** (23) (2004).
58. Z. Lai, F. Peng, H. Wang, H. Yu, S. Zhang and H. Zhao, J. Mater. Chem. A **1** (13), 4182 (2013).
59. T. Ohno, K. Sarukawa and M. Matsumura, New J. Chem. **26** (9), 1167-1170 (2002).
60. C. Gionco, S. Livraghi, S. Maurelli, E. Giamello, S. Tosoni, C. Di Valentin and G. Pacchioni, Chem. Mater. **27** (11), 3936-3945 (2015).
61. S. Sudou, K. Kado and K. Nakamura, Trans. Mater. Res. Soc. Japan **35** (1), 171-174 (2010).
62. A. Y. Choi and C.-H. Han, J. Nanosci.Nanotechnol. **14** (10), 8070-8073 (2014).
63. S. J. Ikhmayies and R. N. Ahmad-Bitar, J. Mater. Res.Technol. **2** (3), 221-227 (2013).
64. A. Paleari, F. Meinardi, A. Lauria, R. Lorenzi, N. Chiodini and S. Brovelli, Appl. Phys. Lett. **91** (14), 141913 (2007).
65. A. S. Hassanien and A. A. Akl, J. Alloys. Compound. **648**, 280-290 (2015).
66. S. Benramache, B. Benhaoua and F. Chabane, J. Semicon. **33** (9), 093001 (2012).